\documentclass[10pt,A4paper,twocolumn,aps,pra, superscriptaddress,longbibliography,nofootinbib]{revtex4-2}

\usepackage{amsmath}
\usepackage{graphicx}
\usepackage{color}
\usepackage[usenames,dvipsnames,svgnames,table]{xcolor}
\usepackage[colorlinks=true, allcolors=blue]{hyperref}
\usepackage{float}
\usepackage{braket}
\usepackage{bbm}
\usepackage{multibib}
\newcites{meth}{Methods References}

\usepackage[capitalise]{cleveref}
\Crefname{figure}{Fig.}{Figs.}
\Crefname{table}{Tab.}{Tabs.}
\usepackage{siunitx}
\usepackage{amssymb}

\usepackage{rotating}
\usepackage{tabularx}
\usepackage{cellspace}
\usepackage{booktabs}
\usepackage{graphics}
\usepackage{newfloat}
\usepackage{soul}

\usepackage{xr}
\usepackage[T1]{fontenc}

\makeatother

\usepackage{chemformula}
\usepackage{microtype}

\begin{document}

\begin{abstract}

Superconducting circuit devices require electrical interconnects between different circuit elements on the chip, for which conventional device architectures use a combination of two structural elements: \textit{airbridges} to connect non-adjacent elements in the base layer, and \textit{bandages} to connect the electrodes forming the Josephson junctions to the base layer. Bandages introduce unwanted parasitic material interfaces and increase the manufacturing complexity. Here, we overcome the limitations imposed by \emph{bandages}  by establishing \textit{all}  electrical interconnects with airbridges of varying size fabricated in a single step. The airbridges show a high yield and mechanical stability over a wide range of sizes from $0.5\,\mu\mathrm{m}$ to $4\,\mu\mathrm{m}$ in width and from $5\,\mu\mathrm{m}$ to $40\,\mu\mathrm{m}$ in length, and show low loss when integrated in coplanar waveguide resonators and transmon qubits. Measured relaxation times up to more than $250\,\mu\mathrm{s}$ in standard transmon geometries show that the process achieves high coherence while substantially easing and accelerating device fabrication.

\end{abstract}

\date{\today}

\author{Prakiran Baidya} 
\affiliation{Department of Physics, Friedrich-Alexander-Universität Erlangen-Nürnberg, Staudtstraße 7, \\91058 Erlangen, Germany}
\affiliation{Quint Computing GmbH, Henkestraße 91, 91052 Erlangen, Germany}

\author{ Momčilo Milosavljević}
\altaffiliation{present address:  i) IMEC,  Kapeldreef 75, 3001 Leuven Belgium ;
 ii) KU Leuven, 3000 Leuven Belgium}
\affiliation{Department of Physics, Friedrich-Alexander-Universität Erlangen-Nürnberg, Staudtstraße 7, \\91058 Erlangen, Germany}

\author{Murali Krishna Kurmapu}
\affiliation{Department of Physics, Friedrich-Alexander-Universität Erlangen-Nürnberg, Staudtstraße 7, \\91058 Erlangen, Germany}

\author{Thomas Fösel}
\affiliation{Department of Physics, Friedrich-Alexander-Universität Erlangen-Nürnberg, Staudtstraße 7, \\91058 Erlangen, Germany}

\author{Harshanth Ram Murugesan}
\affiliation{Department of Physics, Friedrich-Alexander-Universität Erlangen-Nürnberg, Staudtstraße 7, \\91058 Erlangen, Germany}
\affiliation{Quint Computing GmbH, Henkestraße 91, 91052 Erlangen, Germany}

\author{Victor Kemme}
\affiliation{Department of Physics, Friedrich-Alexander-Universität Erlangen-Nürnberg, Staudtstraße 7, \\91058 Erlangen, Germany}

\author{Mojahed Jaber}
\affiliation{Department of Physics, Friedrich-Alexander-Universität Erlangen-Nürnberg, Staudtstraße 7, \\91058 Erlangen, Germany}

\author{Markus Sondermann}
\affiliation{Department of Physics, Friedrich-Alexander-Universität Erlangen-Nürnberg, Staudtstraße 7, \\91058 Erlangen, Germany}

\author{Christopher Eichler}
\email{christopher.eichler@fau.de}
\affiliation{Department of Physics, Friedrich-Alexander-Universität Erlangen-Nürnberg, Staudtstraße 7, \\91058 Erlangen, Germany}
\affiliation{Quint Computing GmbH, Henkestraße 91, 91052 Erlangen, Germany}

\title{Contacting Josephson junctions via airbridges in superconducting circuits}

\maketitle

\section*{Introduction} \label{sec:introduction}
\begin{figure*}[!t]
\centering
	\includegraphics[width=\linewidth]{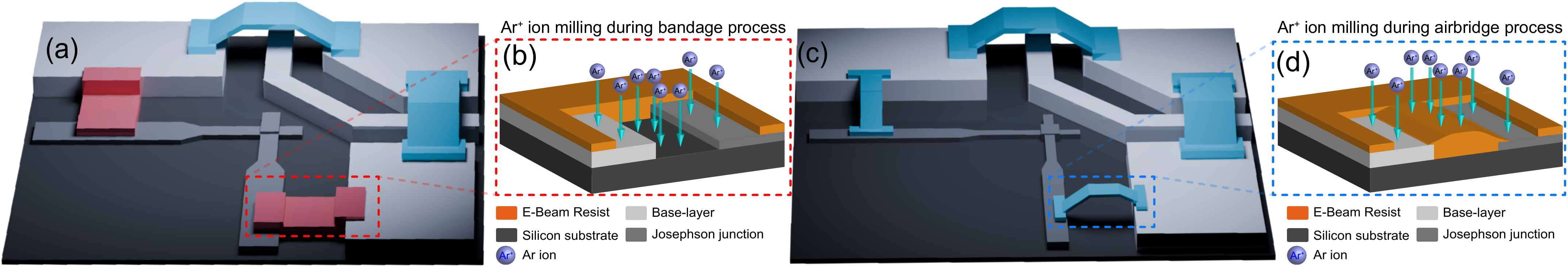}
	\caption{ (a) Schematic of a conventional two-element contacting scheme. Bandages (red) connect Josephson junctions (gray) to the base layer (light gray), whereas airbridges (blue)  connect non-adjacent components of the base layer. (b) Schematic diagram showing the material stack during the in-situ ion milling step applied prior to depositing the bandage metal. (c) Schematic of a the single-element contacting scheme presented in this work, in which all connections are established via airbridges of varying size.  (d) Same as (b) but for the case of  airbridge connections. The bridge scaffold protects the silicon from exposure to ions.}
	\label{fig:fig1} 
\end{figure*}

Superconducting circuits have developed into one of the leading hardware modalities for building quantum information processors \,\cite{Clarke2008,Devoret2013,Blais2021,AbuGhanem2025b} and have enabled a wide range of studies from fundamental quantum physics \cite{Wallraff2004, Hofheinz2008, Gu2017, FornDiaz2019}  to quantum computing \cite{DiCarlo2009, Barends2014, Arute2019}. The sustained progress in scale and quality of superconducting quantum processors has greatly benefited from advancements in the device fabrication based on insights into the underlying material physics \cite{Siddiqi2021, Bland2025, Kurilovich2026}, improvements in the device manufacturing\,\cite{VanDamme2024a, Tuokkola2025} and cleaning\,\cite{Biznarova2024a, ColaoZanuz2025, Lang2026},  and the development of methods for multi-chip integration\,\cite{Conner2021, Norris2024, Rosenberg2017, Grigoras2022, Vahidpour2017a}. Despite all progress, variability in the junction properties together with the presence of material defects in and near the tunnel barrier \cite{Martinis2005,Lisenfeld2019} remain major obstacles to scaling \cite{Mohseni2024}. Furthermore,  the advancements in device engineering fuel a trend toward higher complexity in the device manufacturing, which increases the time from layout to chip. 
\\\\
A typical superconducting quantum device consists of base layer elements, Josephson junctions, and interconnects. Base layer elements have dimensions on the micron-scale and include coplanar waveguides (CPWs), capacitors, and inductors embedded into a common ground plane.  Josephson junctions are formed by overlapping electrodes of sub-micron scale and provide nonlinearity to the circuit dynamics. Interconnects are required to establish electrical connections between different circuit elements. Today's most common planar device architecture uses two types of interconnects for this purpose : free-standing metallic structures called \textit{airbridges} and metallic patches called \textit{bandages}, see Fig.\,\ref{fig:fig1}(a).
\\\\
Airbridges have been widely explored both in classical electronics\,\cite{Dib1991, Simons2001a, Ponchak2005, Pozar2021} and in superconducting quantum devices \,\cite{Wenner2011a, Arute2019}, to suppress crosstalk and to mitigate spurious slotline modes\,\cite{Ponchak2005}. In superconducting devices airbridges are usually  fabricated out of  aluminum \,\cite{chen2014d}. More recently, tantalum \,\cite{Bu2025} and niobium \cite{Bruckmoser2026} have been explored as alternative materials due to their compatibility with acid-based cleaning methods. Having typical lateral dimensions on the order of tens of microns, airbridges are most commonly defined via optical lithography either using single-dose\,\cite{chen2014d}  or using gray-scale exposure \,\cite{Sun2022a, Stavenga2023}. To ease the manufacturing process and enhance flexibility electron-beam lithography has also been explored as an alternative \,\cite{Janzen2022, Fu2026a}.
Previously fabricated airbridges have had sizes predominantly on the order of tens of microns and have therefore been exclusively used to connect non-adjacent elements in the base layer. To date superconducting devices commonly use \emph{bandages} to establish electrical connection between the base layer and the Josephson junction electrodes \,\cite{Dunsworth2017, Nersisyan2019} (see Fig.\,\ref{fig:fig1}(a)). The fabrication of these two types of interconnects --- \emph{airbridges} and \emph{bandages}, require two separate processes,  which increases time, cost, and complexity of manufacturing.  Moreover during conventional bandage processing, the \ch{Si} substrate beneath the junctions suffers unwanted damage from the \ch{Ar+} ion milling used to remove oxides prior to deposition, which degrades device quality\,\cite{Wisbey2010, Bilmes2021b}, as shown in Fig.\,\ref{fig:fig1}(b).  Bandages fabricated in-situ with the Josephson junctions\,\cite{Osman2021, Bilmes2021b} can partly reduce the number of steps involved in the overall chip fabrication workflow, but still face the limitation in the maximal achievable junction size and do not eliminate the risk of \ch{Si} damage as discussed above.
\\\\
In this work, we address these limitations by presenting a unified contacting scheme using a single type
of structural element, see Fig.\,\ref{fig:fig1}(c). More specifically, we employ airbridges of different sizes, all fabricated in a single step using grayscale electron-beam lithography\cite{Janzen2022}.  Apart from reducing the number of fabrication steps we avoid substrate damage in the vicinity of the junctions, since the resist scaffold protects the \ch{Si} substrate from the accelerated \ch{Ar+} ions, as illustrated in Fig.\,\ref{fig:fig1}(d). By consolidating the  fabrication of interconnects, by adding an inherent protection  against the \ch{Ar+} ion damage, and by avoiding parasitic material interfaces, this method provides a path towards high quality quantum devices realized with a significantly leaner process and reduced chemical exposure.
\section{Fabrication and Characterization}\label{sec:characterization}
\begin{figure*}[t]
\centering
	\includegraphics[width=\linewidth]{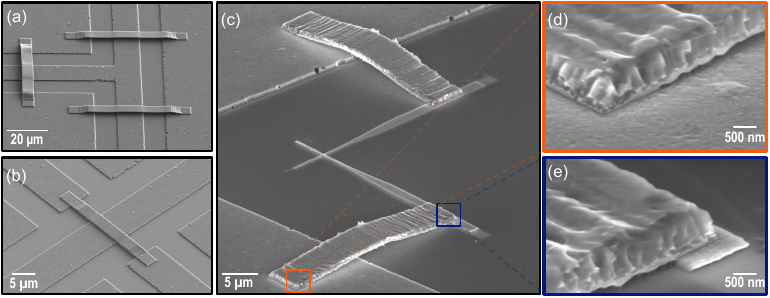}
	\caption{Gallery of scanning electron micrograph (SEM) images of  (a) separate parts of the ground plane contacted via airbridges of length $36\,\mu\mathrm{m}$ and width $4\,\mu\mathrm{m}$, (b) two segments of a coplanar waveguide crossing another one via airbridge of same dimension as in (a), and (c) airbridges having length $15\,\mu\mathrm{m}$ and width $3\,\mu\mathrm{m}$ contact Josephson junction electrodes to the base-layer metal that are $\sim10\,\mu\mathrm{m}$ apart in a test structure. (d)-(e) Zoom into the contact areas between (d) airbridge and base layer and (e) airbridge and Josephson junction electrodes.}
	\label{fig:fig2}
\end{figure*}
Aiming to ease the manufacturing complexity we chose a single-material approach in which we fabricate all superconducting elements from \ch{Al} only.  Following the fabrication of base layer elements and Josephson junction electrodes, we fabricate airbridges using a gray-scale lithography-based process inspired by the work presented in Ref.\,\cite{Janzen2022}, see Appendix\,\ref{sec:supplemental_fab} for details about the fabrication process.  To enable the manufacturing of airbridges of sub-micron size, to allow connections between contact pads of different height, and to enhance mechanical stability and yield, we introduced several modifications relative to the process in Ref.\,\cite{Janzen2022}. Specifically, we reduced the height of the resist threefold from about $3\,\mu$m to $1\,\mu {\mathrm m}$, modified the airbridge design, introduced post-exposure baking, and used a water:IPA solution to develop the resist scaffold  (see Appendices\,\ref{sec:supplemental_design} and\,\ref{sec:supplemental_fab} for details). Using this process, we reliably fabricate airbridges having widths ranging  from about $0.5 \,\mu\mathrm{m}$ to  $4\,\mu\mathrm{m}$ and lengths ranging from  about $5\,\mu\mathrm{m}$ to $40\,\mu\mathrm{m}$, see Fig.\,\ref{fig:fig7} in Appendix\,\ref{sec:supplemental_mechanical}. Access to such a wide range of sizes -- about an order of magnitude both in width and length -- enables the airbridge contacting of sub-micron Josephson junction electrodes to base layer, as well as the fabrication of airbridges across CPWs with cross-sectional dimensions of typically tens of microns, as illustrated in the scanning electron micrographs in Fig.\,\ref{fig:fig2}.
\\\\
All of the airbridges shown in Fig.\,\ref{fig:fig2} were realized in a single grayscale e-beam lithography step within the same airbridge process. The slight bending of the top part of long airbridges across CPW traces, see panels (a) and (b), is due to the non-flat resist scaffold which follows the non-uniform topography of the support structure. The bending above the metal gap areas is particularly pronounced because of the comparably thick base layer of $300\,\mathrm{nm}$ in combination with the comparably thin resist layer of 1$\,\mu$m resulting in a reduced resist height in the etched trenches. Interestingly, this non-uniform bridge profile seems to enhance the structural integrity of the airbridges, allowing for the fabrication of airbridges as long as $\sim60\,\mu\mathrm{m}$ over the CPW lines (see Appendix\,\ref{sec:supplemental_mechanical} for more details). We choose a comparably thick base layer film because it has been shown to reduce susceptibility to interface losses \cite{Biznarova2024a}  and eases marker recognition during the auto-alignment step of  e-beam lithography for patterning the junction-contacting airbridges.  Based on SEM imaging, we consistently find an alignment error of less than $20\,\mathrm{nm}$.  Such high alignment accuracy is essential, as the contacting area on the junction electrode pad measures only $0.8\times3.5\,\mu\mathrm{m}^2$ with the airbridge contact pad dimension being $0.6\times2.5\,\mu\mathrm{m}^2$, and even sub-micron misalignment could result in failed contact or unintended exposure of the \ch{Si} substrate to \ch{Ar+} ions during the milling step. 
\\\\
We systematically study the yield of mechanically stable airbridges and identify the dimension range having the highest yield for device implementation (see Appendix.\,\ref{sec:supplemental_mechanical} for details). We can connect junctions to the base-layer using airbridges with lengths upto $15 \,\mu\mathrm{m}$ as demonstrated by the test structures shown in Fig.\,\ref{fig:fig2}(c). For transmon devices, we use airbridges with a length of  $5\,\mu\mathrm{m}$ and a width of  $3\,\mu\mathrm{m}$ to connect the junctions to the base-layer. Through optical imaging, we inspect $234$ such junction-contacting airbridges across a separate set of $14$ transmon chips fabricated in a single airbridge process and observe only one failed structure due to the breakage of the bridge at the base layer edge.
To connect non-adjacent ground planes across CPW lines we use airbridges with a length of $36\,\mu\mathrm{m}$ and a width of $4\,\mu\mathrm{m}$. We optically inspect a total of $14$ chips containing resonators and transmon devices, with a total of $2000$ grounding airbridges fabricated in a single airbridge process. Among these, we identify approximately $50$ airbridges showing irregularities. Specifically, we observe nine airbridges detached from the base-layer on one side of the contact pad, four airbridges collapsed onto the base-layer and seven airbridges with small metallic protrusions.  The remaining faulty bridges show slight bending across the width but remain suspended without touching the base layer. We attribute these failures to the use of acetone during lift-off. Before dissolving in acetone, PMMA resist initially swells rapidly\,\cite{Janting2019}, which may exert a pulling force on the airbridges, particularly in regions where the bridge staircase overlaps with the development trench (see Appendix\,\ref{sec:supplemental_design} for design detail). We plan to mitigate this effect in the future by using an NMP-based lift-off process. NMP-based solvents have been reported to remove PMMA more effectively than acetone, especially when heated to \SI{80}{\celsius}, enabling  resist dissolution in a more controlled manner\,\cite{gatech_pmma}.
\\\\
Next we verify the electrical connectivity established through the airbridges, by measuring the room-temperature conductance of junctions of varying size. Each junction connects  to one pair of contact pads directly and to another pair via airbridges as shown in Fig.\,\ref{fig:fig10}.  We find the conductance measured through the airbridge-connected contact pads to match the conductance measured through the directly connected contact pads, indicating that the airbridges establish galvanic connections between the elements, see Appendix\,\ref{sec:supplemental_normal_state} for details about experimental setup and data. We further investigated the effect of the airbridge process on the fabricated junctions by measuring the conductance of junctions before and after the airbridge process through directly connected contact pads. We observe a systematic decrease in the measured conductance  of approximately $\sim7\%$ following the process. We attribute this reduction to the soft-baking of the resist at \SI{160}{\celsius} prior to grayscale e-beam lithography exposure (see Appendix\,\ref{sec:supplemental_normal_state} for details). Since the change of junction conductance is reproducible, we can compensate for it during the design.
\section{Electrical Performance}\label{sec:benchmarking}
\subsection{Coplanar Waveguide Resonator Devices} \label{sec:resonator_spectroscopy}
\begin{figure*}[t]
\centering
    \includegraphics[width=\linewidth]{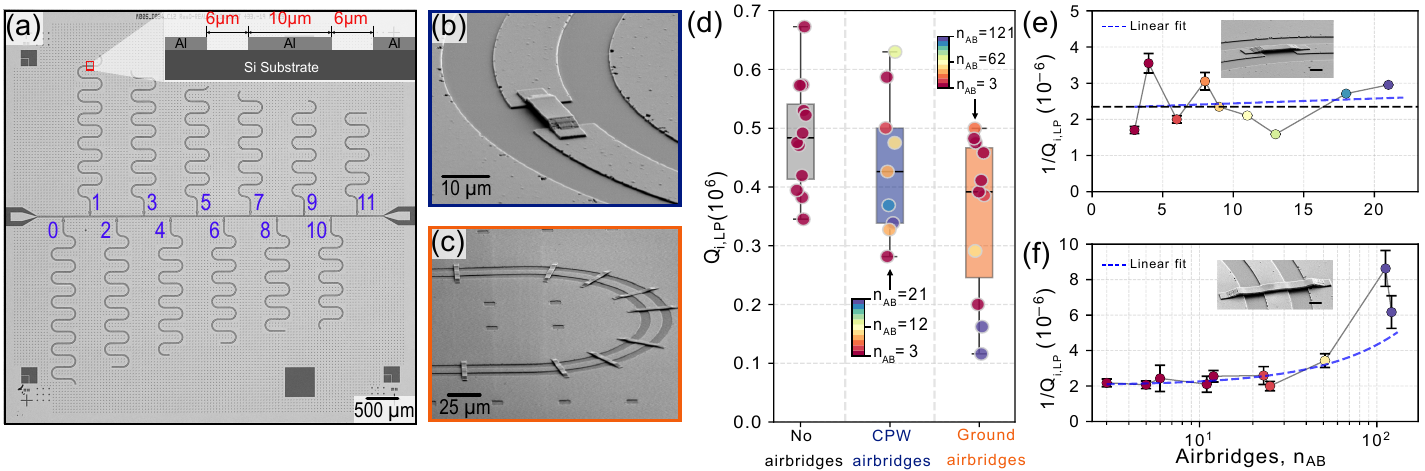}
	\caption{(a) Optical image of one of the three resonator devices each of which contains $12$ CPW resonators of varying length and cross-sectional dimensions as shown in the inset. The numerical label (blue) adjacent to each resonator indicates its index within the chip. (b) SEM image of the second device in which the center conductor is  segmented via connecting airbridges ("CPW airbridges"). (c) Third device with airbridges across the CPW to connect the two halves of the ground plane ("Ground airbridges"). (d) Measured internal quality factors  at approximately single photon level ($Q_{i,\mathrm{LP}}$) for the three devices, with individual data points, median (horizontal bar), and interquartile range (box). The color of each data point indicates the number of airbridges $n_{\mathrm AB}$ contained in the corresponding resonator according to the  colorbar. (e) Measured $1/Q_{i,\mathrm{LP}}$  vs. $n_{AB}$ for  the \emph{CPW-airbridge} device with a linear fit (blue dashed line) and mean value (black horizontal dashed line). (f) Measured $1/Q_{i,\mathrm{LP}}$  vs. $n_{\mathrm{AB}}$ for the \emph{Ground-airbridge}  device with linear fit (blue dashed line). The $n_{AB}$ axis is on log scale for better visibility.}
	\label{fig:fig3}
\end{figure*}
Next, we assess the electrical performance of the fabricated airbridges in the superconducting state at millikelvin temperatures by measuring the quality factor of CPW resonators\,\cite{McRae2020}.  We study three types of resonator devices with different airbridge configurations described below. Each of the devices contains 12 resonators with resonance frequencies equally distributed between $4$ and $8\,\mathrm{GHz}$, set by their respective lengths, see optical image in Fig.\,\ref{fig:fig3}(a). The center conductor of each resonator has a width of $10\,\mu\mathrm{m}$ and is separated  from the adjacent ground planes by a gap of $6\,\mu\mathrm{m}$, see inset of Fig.\,\ref{fig:fig3}(a). On each chip, the resonators are capacitively coupled to a common feedline, with a designed external quality factor $Q_{e}$ of approximately $10^6$. We perform standard transmission spectroscopy measurements at millikelvin temperatures in an experimental setup described in Appendix\,\ref{sec:supplemental_resonator}. We perform a circle-model fit to the data to obtain the resonance frequency and  the internal quality factor as a function of the average intra-resonator photon number for each resonator, see Appendix\,\ref{sec:supplemental_resonator} for details about the methodology.
\\\\
The first of the three chips consists of conventional CPW resonators without any airbridges. The chip has not undergone the airbridge fabrication process and therefore serves as a \emph{reference}. To suppress spurious modes on this chip we connected the two halves of the ground plane using wirebonds (not shown in the picture). For this variant, we measure quality factors in the single-photon regime $Q_{i,\mathrm{LP}}$ on the order of half a million, slightly lower than the state-of-the-art values reported in Ref.\,\cite{Biznarova2024a}  but comparable to typical values reported for \ch{Al}-based CPW resonators with similar geometries \,\cite{chen2014d, Lang2026}.
\\\\
The second chip -- termed \textit{CPW airbridge} -- has segmented central conductors and uses airbridges to connect the individual segments, see Fig.\,\ref{fig:fig3}(b). The number of segments  $n_{\mathrm{AB}}+1$ linearly increases from $4$ to $23$ for the $12$ different resonators. The median $Q_{i,\mathrm{LP}}$ is approximately $4.3\times10^5$ - slightly lower than the median $Q_{i,\mathrm{LP}}$ of the \emph{reference} chip.  Importantly, we find no  significant dependence on the number of airbridges $n_{\mathrm{ AB}}$, which is confirmed by the linear fit (blue dashed line) following closely the mean $\sim2.3\times10^{-6}$ (black dashed line) of  the individual loss values $1/Q_{i,\mathrm{LP}}$ as shown in Fig.\,\ref{fig:fig3}(e). These observations suggest that the airbridges cause no substantial contacting loss and that the slight overall increase in loss is likely due to the additional processing which the chip has undergone during airbridge fabrication. 
\\\\
The third chip -- termed \emph{Ground-airbridge} -- has airbridges crossing over the CPW lines as shown in Fig.\,\ref{fig:fig3}(c) and is otherwise identical to the \textit{reference }chip. The number of these airbridges, $\mathrm{n}_{\mathrm{AB}}$ increases across the 12 resonators from 3 to 121. In stark contrast to the second chip, here, we observe a significant dependence of  $Q_{i,\mathrm{LP}}$ on $n_{\mathrm{AB}}$, suggesting that the airbridges across the CPW make an additional loss contribution.  To estimate the loss per airbridge we perform a linear fit to the internal loss $1/Q_{i,\mathrm{LP}}$ vs. $\mathrm{n}_{\mathrm{AB}}$ which yields an estimated average loss per airbridge of $\sim2.3\times10^{-8}$ as shown in Fig.\,\ref{fig:fig3}(f). The fit matches well till $n_{\mathrm{AB}} \approx 50$ supporting the conjecture of linear scaling between loss and the number of airbridges. Moreover, the intercept of the linear fit, which represents the loss in the absence of airbridges, gives $1/Q_{i,\mathrm{LP}}$ of $\sim2.03\times10^{-6}$. This value matches well with the median loss $1/Q_{i,\mathrm{LP}}\sim2.07\times10^{-6}$  calculated from the $Q_{i,\mathrm{LP}}$ values of the \emph{reference} chip shown in Fig.\,\ref{fig:fig3}(d). The noticeable deviation between the fit and individual data points for $\mathrm{n}_{\mathrm{AB}}\ge100$, we attribute to the overall variations of $Q_{i,\mathrm{LP}}$ observed in the corresponding data in Fig.\,\ref{fig:fig3}(d).  The additional loss associated with grounding airbridges suggests the presence of lossy dielectrics, which could stem from resist residues beneath the airbridges not having been fully removed during lift-off. Although measurable, the average dielectric loss per airbridge extracted from the linear fit, remains two orders of magnitude lower than the intrinsic resonator loss, showing that the airbridges serve as high-quality interconnects in superconducting quantum devices.  
Furthermore, we see great potential for further process optimization, e.g. by using heated NMP-based solvents instead of acetone for the lift-off (Appendix\,\ref{sec:supplemental_fab}) and by applying additional cleaning steps after fabrication.  
\\\\      
\subsection{Transmon devices}\label{sec: qubit_spectroscopy}
\begin{figure}[t!]
\centering
\includegraphics[width=\linewidth]{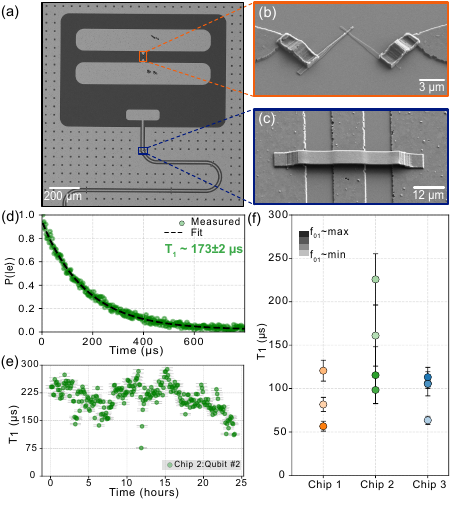}
\caption{(a) Optical image of a two-pad transmon device with a design similar to the one used in Ref.\,\cite{Bland2025}. (b)-(c) SEM image of the section marked in (a) showing the airbridges connecting two junction electrodes to the base layer in (b), and an airbridge crossing a CPW in (c). (d) Measured qubit population (green circle) as a function of delay time for one of the transmon devices after an intial $\pi$-pulse. The dashed line is a fit to the data. (e) Repeated $T_1$  measurements performed over $24\,\mathrm{hours}$ for the best performing qubit. (f) Measured  $T_1$ for four different qubits across three chips. The color saturation of data points indicates the qubit frequency $f_{01}$ (see  color bar).}
\label{fig:fig4}
\end{figure}
We further verify the electrical performance of airbridges when used for contacting to Josephson junctions by fabricating fixed-frequency floating transmon devices, see Fig.\,\ref{fig:fig4}(a) .  Importantly, both the airbridges used for junction-contacting and for grounding, see Fig.\,\ref{fig:fig4}(b) and (c) respectively, are fabricated within a single airbridge process. We choose a design with large capacitor pads \,\cite{Bland2025} and with a comparably weak coupling to the readout circuitry, compare Table\,\ref{tab:transmon_parameter} in the Appendix\,\ref{sec:supplemental_transmon}, in order to minimize energy relaxation caused by sources not directly related to the Josephson junctions and the elements in their near vicinity. Each chip contains four transmon qubits with transition frequencies ranging from $4.3\,\mathrm{GHz}$ to $5.3\,\mathrm{GHz}$, and with a charging energy $\mathrm{E}_c/2\pi$ of $208\,\mathrm{MHz}$. Each transmon is dispersively coupled to a dedicated resonator with a resonance frequency between $6\,\mathrm{GHz}$ and $7\,\mathrm{GHz}$. The readout resonators couple to a common feedline via dedicated Purcell filters, used for control and readout, see Appendix\,\ref{sec:supplemental_transmon} for details about the experimental setup and the device parameters. 
\\\\
We determine the energy relaxation times $T_1$ for the qubits of three identically designed chips via time-domain measurements. 
We obtain $T_1$ by fitting an exponential decay model to the excited state population measured after an initial $\pi$-pulse and a variable delay time as shown in Fig.\,\ref{fig:fig4}(d) for one of the transmon devices. For the best performing transmon among the three chips, the fit yields a peak $T_1$ of $\sim290\,\mu\mathrm{s}$. To investigate the stability of $T_1$ over long timescales, we perform $300$ independent repeated time-domain measurements of the same transmon over a total duration of  $24$\,hours, as shown in Fig.\,\ref{fig:fig4}(e). 
We observe temporal fluctuations with a standard deviation $\sigma$ of $\sim29\,\mu\mathrm{s}$ and a distribution that is well described by a Gaussian centered at $\sim225\,\mu\mathrm{s}$, as shown in Fig.\,\ref{fig:fig15}(d) in Appendix\,\ref{sec:supplemental_transmon}. Such temporal fluctuations are typical for transmon devices and are commonly attributed to the slow dynamics of  the two-level systems that the transmons couple to\,\cite{Schloer2019, Burnett2019}.
\\\\
We further perform repeated $T_1$ measurements of all the transmons across the three chips and extract the median value for each qubit. The resulting $T_1$ distribution, shown in Fig.\,\ref{fig:fig4}(f), indicates median $T_1$ values around $100\,\mu\mathrm{s}$ for the majority of the transmons. Except for a few outliers, the measured  $T_1$  are comparable to the state-of-the-art for all-\ch{Al}  transmon devices\,\cite{Biznarova2024a}, highlighting the potential of the presented airbridge process to manufacture high-quality superconducting circuit devices with a streamlined process. 
\section*{Conclusion and Outlook}\label{sec:conclusion}
Here we have demonstrated an airbridge fabrication process based on a single-step gray-scale electron-beam lithography technique that unifies the formation of electrical interconnects. We implemented this process in an all-aluminum device architecture. However, the process is fully compatible also with other base-layer 
materials such as Niobium and Tantalum.
\\\\
We studied the electrical performance of the  airbridges by measuring both the internal quality factor of CPW resonators and the energy relaxation times of transmon devices. Our results indicate that airbridges introduce no measurable contacting loss and only minimal dielectric loss. We anticipate further improvements by optimizing the lift-off process for example by using heated NMP-based solvents and by introducing post-fabrication cleaning steps such as plasma ashing, ozone cleaning, acid dipping or surface passivation prior to cooldown\,\cite{Marcaud2025, Lang2026}.
\\\\
In summary, our results highlight that the demonstrated process yields robust and scalable interconnects for superconducting devices, particularly well suited in research and development settings requiring fast turn-around times and resource-efficient manufacturing. Moreover, the wide dimensional range achieved with this process enables the use of these airbridges to realize more complex device geometries such as gradiometric SQUIDs and on-chip spiral inductors.
\section*{Acknowledgments}
The authors thank Jean-Claude Besse for valuable discussions and feedback on the manuscript. The authors acknowledge financial support by the Bavarian StMWK through the MQV lighthouse project QuMeCo and the accelerator project E$^3$QC, by the German Federal BMFTR through the project MuniQC-SC, by the German Research Foundation (DFG) through the grants INST 90/1430-1 FUGG, INST 90/1435-1 FUGG, INST 90/1439-1 FUGG and by FAU. The authors acknowledge support by Irina Harder, Olga Ohletz, Alexander Gumann, Katrin Ludwig, and Florentina Gannott from the TDSU Micro and Nanostructuring unit at the Max Planck Institute for the Science of Light, where the majority of the device fabrication has been carried out.
\section*{Author contributions}
M.M. and P.B. developed the airbridge fabrication process and performed AFM and SEM characterization. P.B. designed the resonator and transmon devices. P.B., M.K.K., V.K fabricated the devices. T.F., H.R.M., M.K.K.  contributed to the measurement protocol. M.J., T.F., M.S. contributed to the cryogenic measurement system.  P.B., M.K.K. acquired and analyzed the data. P.B. and C.E. wrote the manuscript with input from all authors. C.E. conceived and supervised the project. 


\begin{appendix}

\section{Air bridge design}\label{sec:supplemental_design}

\begin{figure}[t]
\centering
	\includegraphics[width=\linewidth]{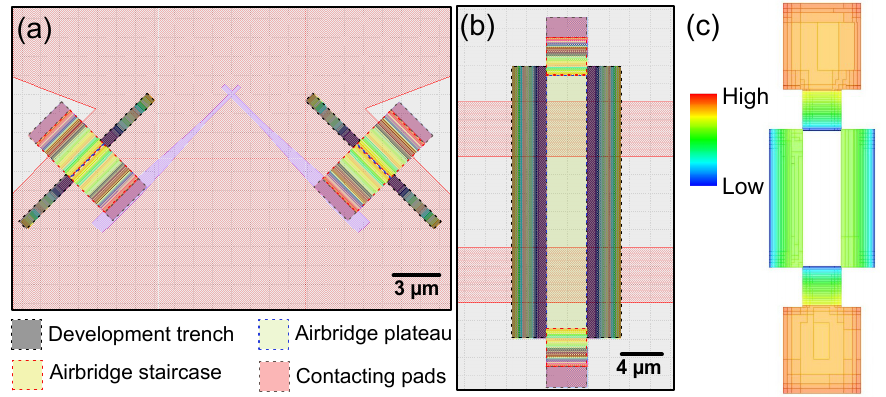}
	\caption{GDS layout of a typical airbridge design used for contacting -- (a) junction to the base-layer, and (b) two non-adjacent ground planes across a CPW . For both panels (a) and (b), the shaded regions designate different parts of the GDS design as indicated. (c) Dose distribution used to define the gradient exposure during gray-scale e-beam lithography process. }
	\label{fig:fig5} 		
\end{figure}
We use a design for the airbridges similar to the ones used in \,\cite{Papageorgiou2013, Girgis2006, Janzen2022}. The design consists of two contacting pads,  two staircase regions and a plateau region with two adjacent development trenches, as illustrated and marked in Fig.\,\ref{fig:fig5}(a)-(b).
We define the height profile by $28$ discrete layers to each of which we assign a corresponding exposure dose during gray-scale e-beam lithography, as visualized in Fig.\,\ref{fig:fig5}(c). We assign the clearance dose to the contact pad region and gradually decrease the dose from layer to layer until we reach zero dose in the plateau region. We include the development trench in the design to disconnect the metal film deposited on top of the unexposed resist area from the deposited metal on top of the plateau region to facilitate the lift-off. To ensure that the contact pad of the airbridge makes contact only with the junction metal but not with the \ch{Si} we chose the width of the junction lead to be 800$\,$nm -- about 200$\,$nm wider than the base of the bridge.  
\begin{figure*}[t]
\centering
    \includegraphics[width=\linewidth]{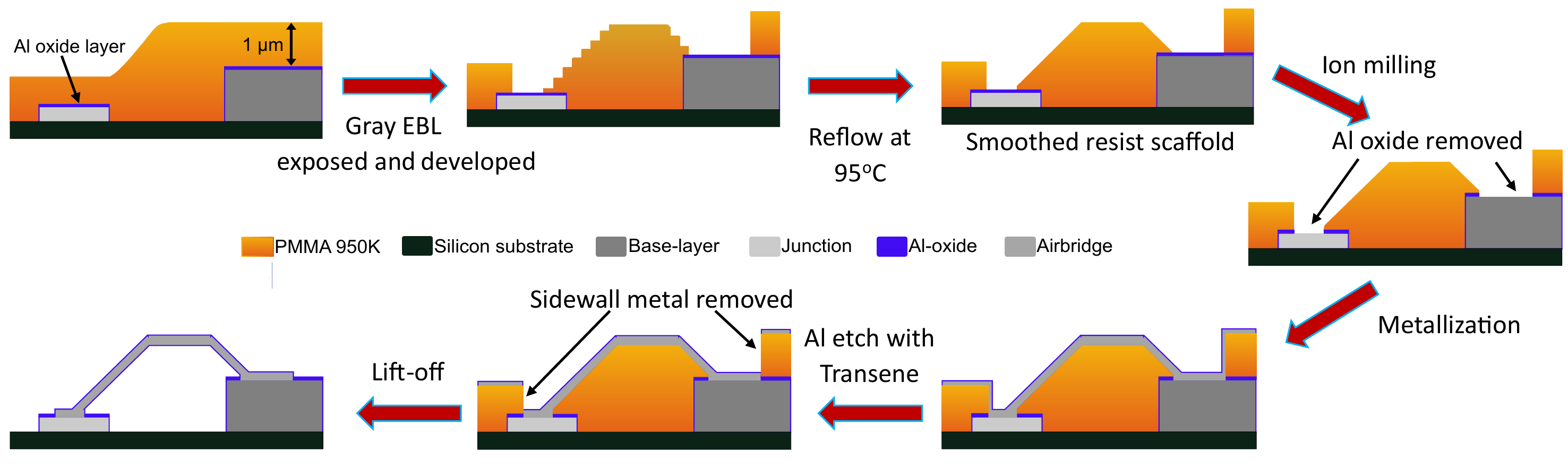}
	\caption{Schematics of the fabrication process flow for the airbridge structures (not to scale).}
	\label{fig:fig6}
\end{figure*}
\section{Device fabrication}
\label{sec:supplemental_fab}
We start the device fabrication process by dipping a high-resistivity  4-inch silicon wafer ($>10\,\mathrm{k}\Omega\mathrm{cm}$) from Siegert Wafer into a 5\%  HF solution to remove the native \ch{SiO2} layer. Immediately after, we transfer the wafer to an e-beam evaporator tool (Quantum Series from Angstrom Engineering). We heat the wafer to \SI{300}{\celsius} prior to the deposition of a \SI{300}{nm} thin film of Al at a rate of  \SI{2}{nm/s}. Before unloading the sample from the vacuum chamber, we cap the deposited metal by an in-situ oxidation achieved by keeping the wafer for 10 min in the chamber filled with $20\,\mathrm{mbar}$ of \ch{O2}.  We then pattern the base-layer using a standard photolithography process followed by development and  etching in a AZ 726MIF solution. We then coat the wafer with AZ P4K  as a dicing  protection layer and dice it into $30\times30\,\mathrm{mm}^2$ blocks.  We finally strip off the protective coating block-wise using a wet chemical process to prepare the sample for subsequent fabrication steps. 
\\\\
To fabricate the Josephson junctions we use a bilayer of e-beam resist.  We spin-coat $\sim1000\,\mathrm{nm}$ of  MMA as a bottom layer baked at \SI{150}{\celsius} for $90\,\mathrm{s}$ and $\sim150\,\mathrm{nm}$ of  ZEP520A (1:1 anisole diluted solution) baked at \SI{180}{\celsius} for $180\,\mathrm{s}$ as a top layer. We define a Manhattan style Josephson junction pattern using a \SI{100}{keV} e-beam lithography system (EBPG5200 from Raith GmbH) with the dose of $1100\,\mu\mathrm{C}/\mathrm{cm}^2$ and $198\,\mu\mathrm{C}/\mathrm{cm}^2$ for main and overhang feature respectively.  After exposure, we perform a two-step development to define the junction pattern. First, we develop in ZEDN50 (n-amyl acetate) at \SI{0}{\celsius} for $1\,\mathrm{min}$ followed by a $1\,\mathrm{min}$ stopping in \ch{IPA}. Second, we develop for $5\,\mathrm{min}$ in MIBK:IPA (1:3)  followed by a  $30\,\mathrm{s}$  stopping in \ch{IPA}. Before proceeding with the metallization, we use a mild ashing process in a RIE-F etching tool (Fluor RIE/ICP from Oxford Instruments) using \ch{O2} plasma for $45\,\mathrm{s}$ at $6\times10^{-3}\,\mathrm{mbar}$ pressure with $150\,\mathrm{W}$ RF power to remove any residual resist from the exposed area.  We then load the sample into the deposition tool, evacuate the loadlock, and keep the sample overnight before proceeding.  We then transfer the sample into the deposition chamber which has a pressure below $2\times10^{-8}\,\mathrm{mbar}$, rotate the sample to the correct planetary angle, tilt the sample stage to \SI{45}{\deg}  and deposit  $20\,\mathrm{nm}$ of \ch{Al} at a deposition rate of $0.5\,\mathrm{nm}/\mathrm{s}$. We define the oxide barrier by  controllably oxidizing the deposited \ch{Al} film  in a mixture of  $25\%$ of \ch{O2} in \ch{Ar} gas.  We then proceed with depositing the top electrode. For this we deposit $100\,\mathrm{nm}$ of \ch{Al} two times in two separate planetary angle positions - one at \SI{90}{\deg} with respect to the bottom electrode angle position and the second one at \SI{180}{deg} relative to the first one. This ensures that the top electrode overlaps the bottom one from both sides at the junction region. Before unloading, we cap the \ch{Al} film by in-situ oxidation as mentioned earlier. Afterwards we unload the sample and proceed with the lift-off process in a sequence of  $30\,\mathrm{min}$ of  acetone at \SI{50}{\celsius}, followed by $5\,\mathrm{min}$ of Dimethyl Sulfoxide (DMSO) at \SI{80}{\celsius} and sonication for $5\,\mathrm{min}$, followed by sonication in acetone kept at \SI{50}{\celsius} for $5\,\mathrm{min}$, rinsing with IPA and blow-dry with \ch{N2} gas.
\\\\
The airbridge fabrication process is schematically shown in Fig. \,\ref{fig:fig6}. We start by spin-coating $\sim1000\,\mathrm{nm}$ of  PMMA950K e-beam resist and soft-bake at \SI{160}{\celsius} for  $5\,\mathrm{min}$. To define the airbridge structure we pattern the resist layer with grayscale e-beam lithography.  An important aspect of the grayscale exposure is to calibrate the electron dose to the height of the resist removed during development. This dose-height calibration is estimated using the contrast curve measurement. Before the exposure,  we assign the appropriate dose to the different layers in the design by implementing proximity error correction and the contrast curve calibration through the BEAMER software from GenISys.  We then expose the sample in a $100\,\mathrm{keV}$ EBL system with a base dose of $350\,\mu\mathrm{C}/\mathrm{cm}^2$. Before moving forward with the development, we do a post-exposure baking at \SI{80}{\celsius} for $3\,\mathrm{min}$,  as we noticed that the time between exposure and development influences the contrast of the resist\,\cite{Mortelmans2020}. We develop the sample in Water:IPA(3:7) solution for $95\,\mathrm{s}$. The choice of the developer solution over the more conventional one such as  MIBK:IPA\,\cite{Janzen2022} has one of the main advantages that the slope of the contrast curve in this case is lower, which is beneficial for grayscale lithography. Moreover, for the MIBK:IPA solution the resist becomes more susceptible to cracking post-development specifically for thicker resist stack which we avoid with our choice of developer solution. We afterwards perform reflow of the resist stack at \SI{100}{\celsius} for $5\,\mathrm{min}$. We then proceed with the deposition step, which starts with an in-situ \ch{Ar} ion milling process within the deposition tool for $150\,\mathrm{s}$ to etch off the oxide layer from the airbridge contact pad region. We deposit $450\,\mathrm{nm}$ of \ch{Al} at a deposition rate of $0.5\,\mathrm{nm}/\mathrm{s}$. Similar to the earlier two processes, before unloading from the deposition tool we cap the \ch{Al} film by in-situ oxidation as mentioned earlier. Before lift-off, we dip the sample in Transene Al etchant for $70\,\mathrm{s}$ to etch $\sim75\,\mathrm{nm}$ of \ch{Al}. This step  removes the \ch{Al} film deposited on the resist sidewall, which in turn facilitates the lift-off. We perform lift-off  in the following sequence : keeping the sample in acetone overnight at room-temperature, followed by $15\,\mathrm{min}$ in DMSO at \SI{80}{\celsius}, $10\,\mathrm{min}$ in acetone at \SI{50}{\celsius}, rinsing with IPA and blow-dry with \ch{N2} gas. During drying with the \ch{N2} we make sure that the gas is not being applied to the sample surface at high pressure, which can lead to damaged airbridge structures\,\cite{Bolgar2025}. We subsequently coat the processed $30\times30\,\mathrm{mm}^2$  block with a bi-layer coating of $\sim1\,\mu\mathrm{m}$ of MMA and $\sim5\,\mu\mathrm{m}$ of AZ P4K as the dicing protection layer and dice it into $7\times7\,\mathrm{mm}^2$ chips. We clean these chips in sequence of $15\,\mathrm{min}$ of DMSO at \SI{80}{\celsius}, $5\,\mathrm{min}$ of acetone at \SI{50}{\celsius}, IPA rinse and mild blow dry with \ch{N2} gas.
\section{Mechanical stability of air bridges}
\label{sec:supplemental_mechanical}
\begin{figure}[t]
\centering
	\includegraphics[width=\linewidth]{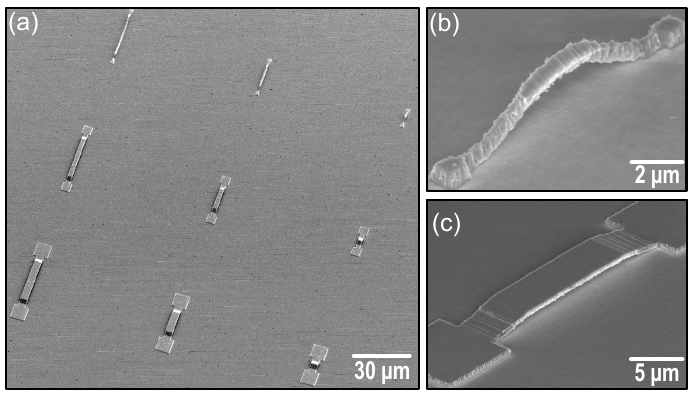}
	\caption{SEM image of -- (a) arrays of airbridges with varying width and length, (b) the narrowest stable airbridge structure with a width of $500\,\mathrm{nm}$ that can be fabricated using the current process, (c) a stable $\sim30\,\mu\mathrm{m}$ long airbridge structure that is ideal to connect ground planes across CPW. }
	\label{fig:fig7}
\end{figure}
To identify the dimension range of mechanically stable airbridges, we fabricate arrays of airbridges of varying length, width and slope on top of an \ch{Al} film deposited on a \ch{Si} substrate, as shown in Fig.\,\ref{fig:fig7}(a). We fabricate airbridges down to a  width of $500\,\mathrm{nm}$ and $8\,\mu\mathrm{m}$ in length as shown in Fig.\,\ref{fig:fig7}(b),  setting the low-dimensional limit of mechanically stable airbridges that can be fabricated using the process. However, the yield of such narrow airbridges are $<80\%$. Through optical micrograph inspection we find that the airbridges having the width ranging from $2\,\mu\mathrm{m}$ to $8.5\,\mu\mathrm{m}$ and the length ranging from $5\,\mu\mathrm{m}$ to $\sim40\,\mu\mathrm{m}$ [Fig.\,\ref{fig:fig7}(c)] have the highest survival rate with $>99\%$ yield. Through optical characterization of the yield we choose airbridges having width of $4\,\mu\mathrm{m}$, staircase length of $4.14\,\mu\mathrm{m}$ and contact pad length of $2.25\,\mu\mathrm{m}$ to connect non-adjacent ground planes.  To connect the junction to the base layer we choose airbridge structures of width $2.5\,\mu\mathrm{m}$, staircase length $2.5\,\mu\mathrm{m}$ and contact pad length $0.6\,\mu\mathrm{m}$. 
\begin{figure}[t]
\centering
	\includegraphics[width=\linewidth]{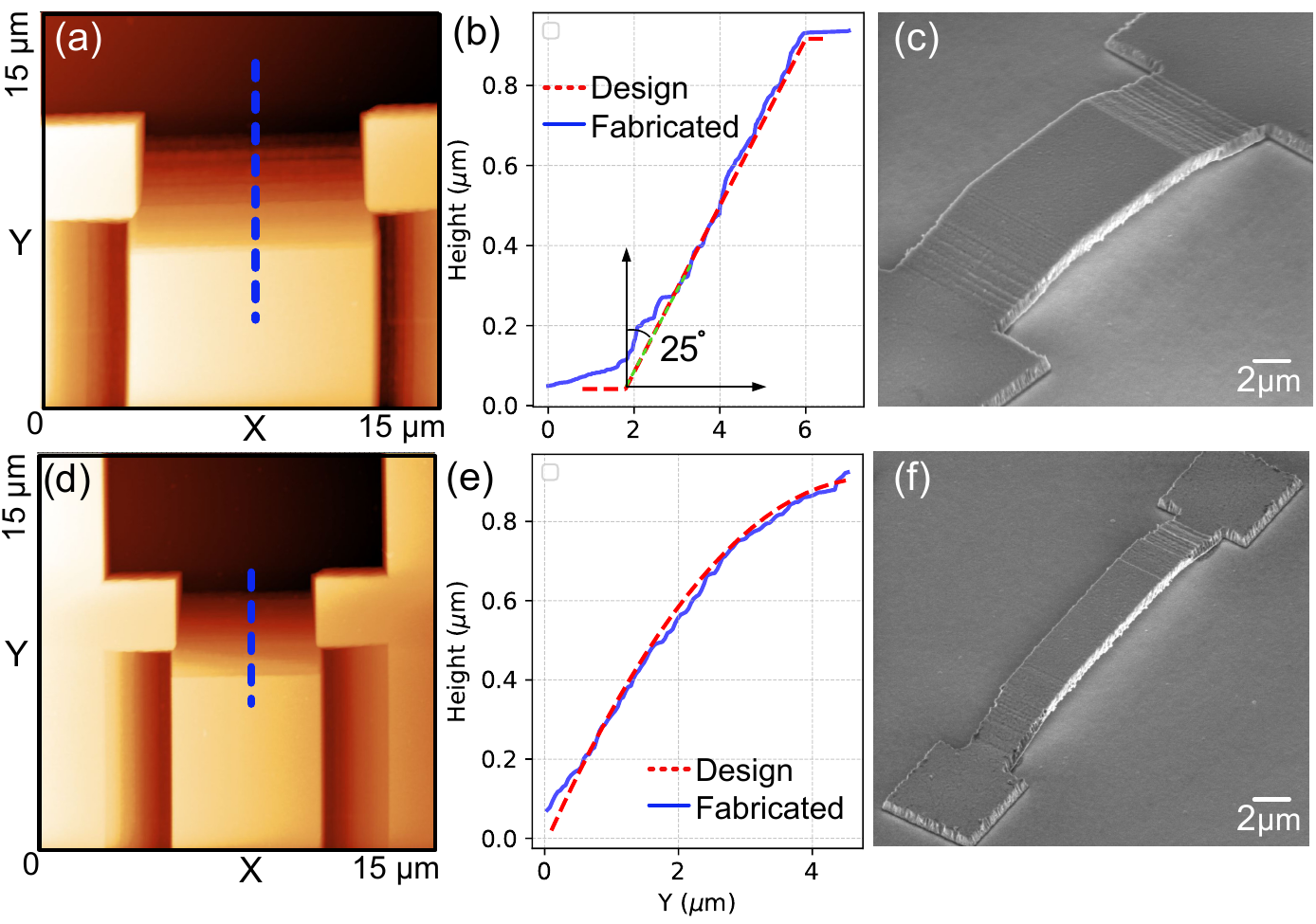}
	\caption{Analysis of linear (a-c) and parabolic (d-f) air bridge staircase profiles: 
(a,d) AFM image of the resist profile in the staircase region; 
(b,e) height profile along the dashed line drawn in the AFM images (blue solid line) and profile fit (red dotted line); 
(c,f) SEM image of the fabricated air bridge.}
	\label{fig:fig8} 	
\end{figure}
 \\\\
The profile of the airbridge staircase region largely influences the stability of the structure.  We tested linear and parabolic functions for this purpose [Fig.\,\ref{fig:fig8}] as indicated by the height plots shown in Fig.\,\ref{fig:fig8}(b) and (e). 
Using the yield tests similar to the one mentioned earlier, we find that the linear staircase design produces mechanically stable airbridges over a wider dimensional range than the parabolic design, with a yield exceeding $99\%$. We thereafter implemented the linear staircase profile as the standard for subsequent process development. For the process, we choose a staircase angle of $\sim25$ degrees as shown in Fig.\,\ref{fig:fig8}(b), to ensure sufficient thickness of metal is deposited on the inclined staircase region and thereby maintain the structural integrity of the airbridges. 
\\\\
As mentioned in the main text, we observe bending of the plateau region for those airbridges bridging the gap over a CPW line. To study the implications on the structural integrity we fabricate airbridges of same length  ($\sim36\,\mu\mathrm{m}$) to connect non-adjacent ground planes having two different topographies. In the first case the airbridge crosses over a CPW line, see Fig.\,\ref{fig:fig9}(a). In the second case the airbridge crosses over a $20\,\mu\mathrm{m}$ wide trench, see Fig.\,\ref{fig:fig9}(b). In the majority of instances we observe, the structure in Fig.\,\ref{fig:fig9} (a) is stable but the other one failed as shown in Fig. \,\ref{fig:fig9}(b). We attribute this stability to the presence of the CPW line in the middle, that adds a curvilinear profile to the airbridge plateau region, which is in contrast to the designed horizontal profile indicated by the implemented dose distribution in Fig.\,\ref{fig:fig5}(c).  Similar to a recent report in which airbridge stability has been enhanced through engineering the bridge geometry\,\cite{Bolgar2025}, the intrinsic modulation of the CPW airbridges in the plateau region in our case further reinforces the overall structure and extends the achievable length range for mechanically stable airbridges as mentioned in the main text in Sec.\,\ref{sec:characterization}.
\begin{figure}[t]
\centering
\includegraphics[width=\linewidth]{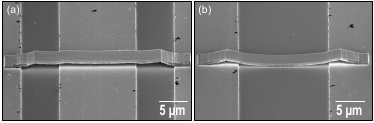}
	\caption{SEM image of -- (a) a stable airbridge having a length of $36\,\mu\mathrm{m}$ that was fabricated to connect two non-adjacent ground planes across a CPW central conductor. The ground planes are $32\,\mu\mathrm{m}$ apart.  (b) A failed airbridge structure of the same length as in (a) but without any central conductor in the middle, connecting two non-adjacent ground plane segments that are $20\,\mu\mathrm{m}$ apart. }
	\label{fig:fig9}
\end{figure}
\section{Normal state resistance}
\label{sec:supplemental_normal_state}
\begin{figure}[t]
   \centering
    \includegraphics[width=\linewidth]{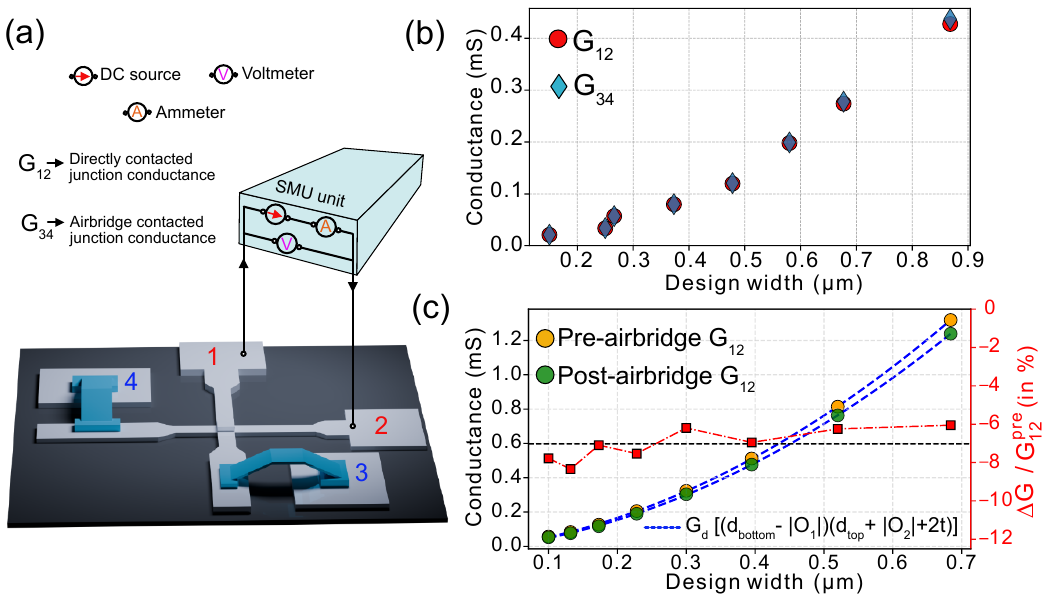}
	\caption{(a) Schematic of the device structure and measurement setup used for room-temperature electrical characterization. Single junction is connected to measurement contact pads both directly ("Directly contacted") and via airbridges ("Airbridge contacted") with measured room-temperature  conductance denoted as $\mathrm{G}_{12}$ for directly contacted  and $\mathrm{G}_{34}$  for airbridge contacted. (b) Measured room-temperature conductance for junctions with different design widths in both the configurations as mentioned in (a). (c) Measured  room-temperature conductance for directly contacted junctions with different design widths before (orange circle) and after (green circle) the airbridge process, shown on the black left axis. The blue dashed line shows a fit to the junction conductance equation in the inset. The corresponding percentage change in conductance $\Delta G /G_{12}^{\mathrm{pre}} = \left(G_{12}^{\mathrm{post}}-G_{12}^{\mathrm{pre}}\right)/G_{12}^{\mathrm{pre}}$ shown on the red right axis with the black dashed line indicating the mean change of $\sim7\%$.  
    }
	\label{fig:fig10} 
\end{figure}

We benchmark the quality of the electrical contacts established by the airbridges through room-temperature conductance measurements. We contact Josephson junctions via two different configurations as mentioned in the main text in Sec.\,\ref{sec:characterization} and as shown in Fig.\,\ref{fig:fig10}(a). From the plot of the measured conductance values for both  configurations, i.e. directly contacted $G_{12}$ and airbridge contacted $G_{34}$ as shown in Fig.\,\ref{fig:fig10}(b), we can infer that the fabricated airbridges establish galvanic contact between the elements. 
\\\\
Additionally, we observe an overall decrease in the conductance by $\sim7\%$  after the airbridge process, see Fig.\,\ref{fig:fig10}(c), which we attribute to the soft-baking during the resist coating step required for e-beam lithography process\,\cite{Migacz2003}. Similar observations have been reported previously in Ref.\,\cite{Janzen2022}. In our case, the effect is less pronounced than in Ref.\,\cite{Janzen2022}, possibly due to the use of a lower baking temperature  (\SI{160}{\celsius}).
\begin{figure*}[t]
\centering
    \includegraphics[width=0.65\linewidth]{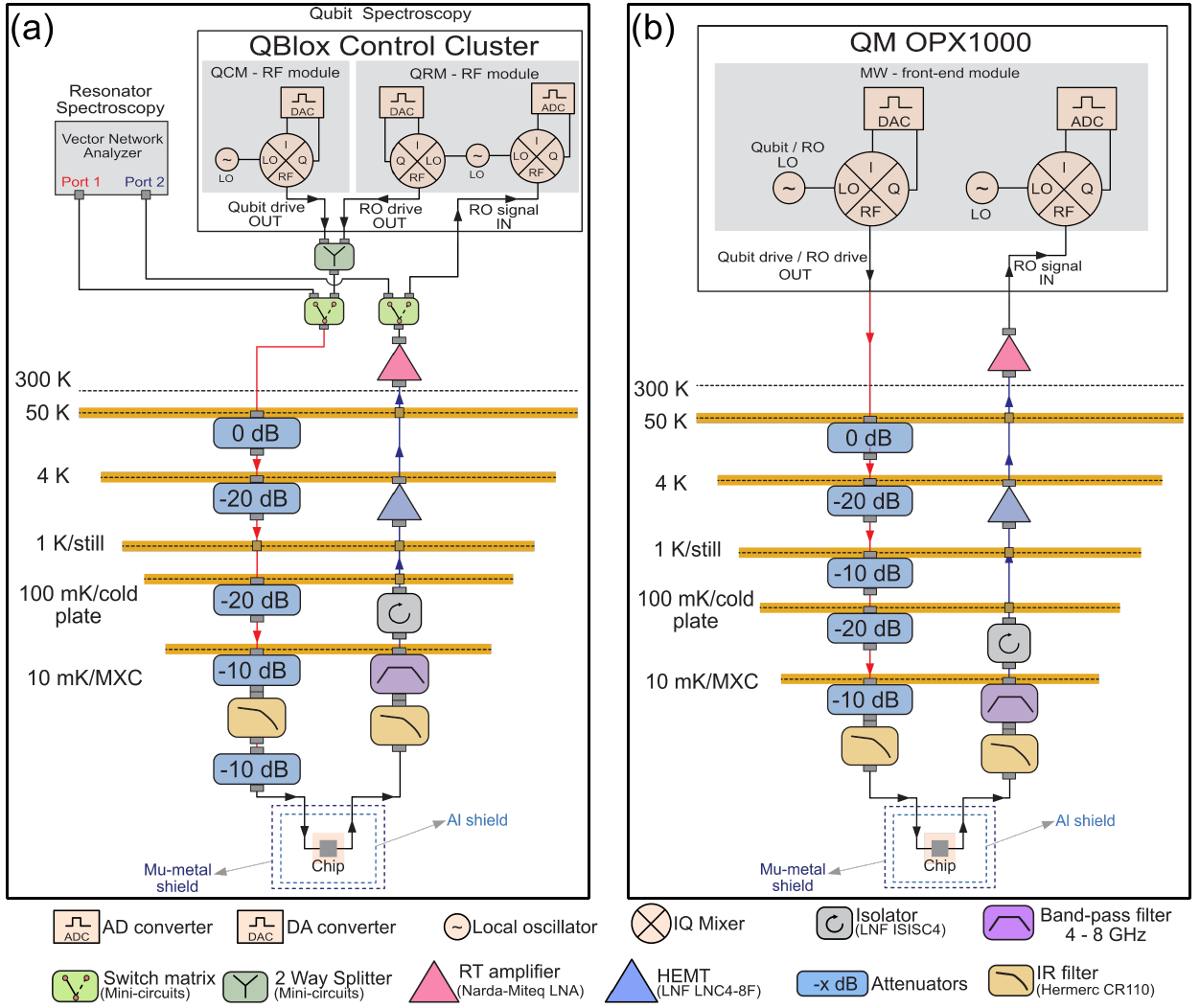}
	\caption{Wiring diagram of - (a) \emph{Cryogenic Measurement Setup 1} used for resonator and qubit spectroscopy, (b) \emph{Cryogenic Measurement Setup 2} used for qubit spectroscopy. 
    }
	\label{fig:fig11}
\end{figure*}
\section{CPW resonator measurement}
\label{sec:supplemental_resonator}
\begin{figure}[t]
\centering
    \includegraphics[width=\linewidth]{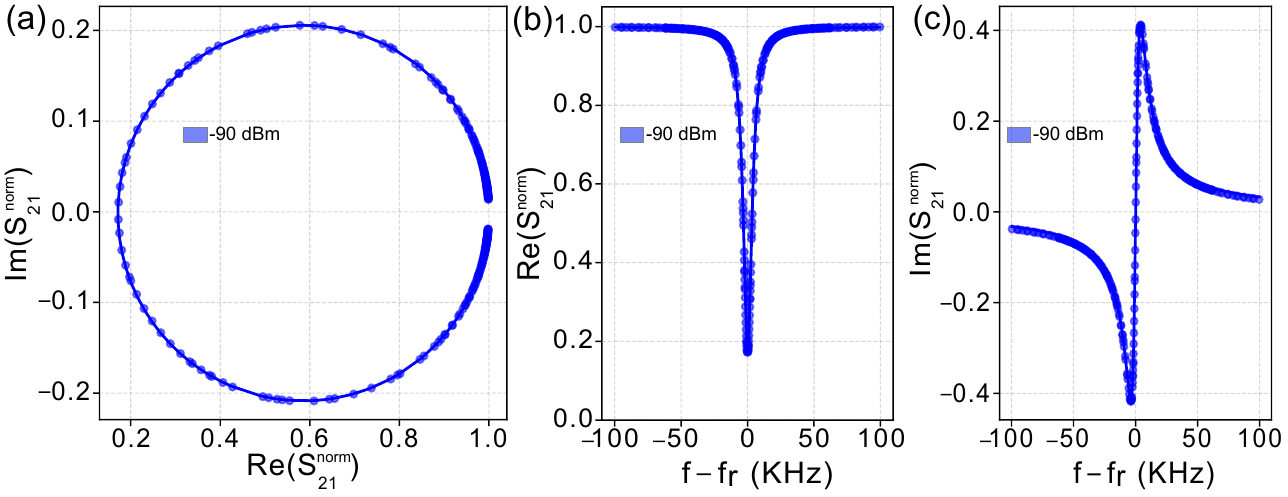}
	\caption{Normalized $S_{21}$ data (blue circles) along with the fit (blue solid line) for one of the resonators plotted in - (a) the complex plane, (b) the real plane as function of frequency, (c) the imaginary plane as a function of frequency. $\mathrm{f}_\mathrm{r}$ in panel (b) and (c ) denotes the resonance frequency of the corresponding resonator. }
	\label{fig:fig12}
\end{figure}
\begin{figure}[t]
\centering
    \includegraphics[width=\linewidth]{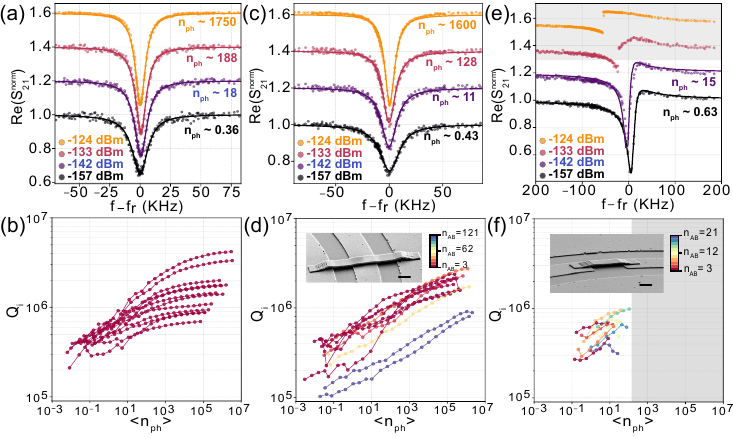}
	\caption{Real normalized part of the measured transmission spectra  $S_{21}^{\mathrm{norm}}$ (circles) along with the fit (solid line) as function of varying input power $P_{in}$ for a resonator on \emph{reference} chip in (a), on \emph{Ground-airbridge} chip in (c), on \emph{CPW-airbridge} chip in (e). The plots for different powers are shifted vertically for visualization. Evaluated internal quality factor $Q_{i}$ of all the resonators as a function of average photon number $\left<n_{ph}\right>$ for \emph{reference} chip in (b), for \emph{Ground-airbridge} chip in (d), for \emph{CPW-airbridge} chip in (f). The colorbar in (d) and (f) shows corresponding airbridge number on each resonator. The gray region in (e) and (f) indicates the high power regime where the linear model fit fails. 
    }
	\label{fig:fig13}
\end{figure}
\begin{figure}[t]
\centering
    \includegraphics[width=\linewidth]{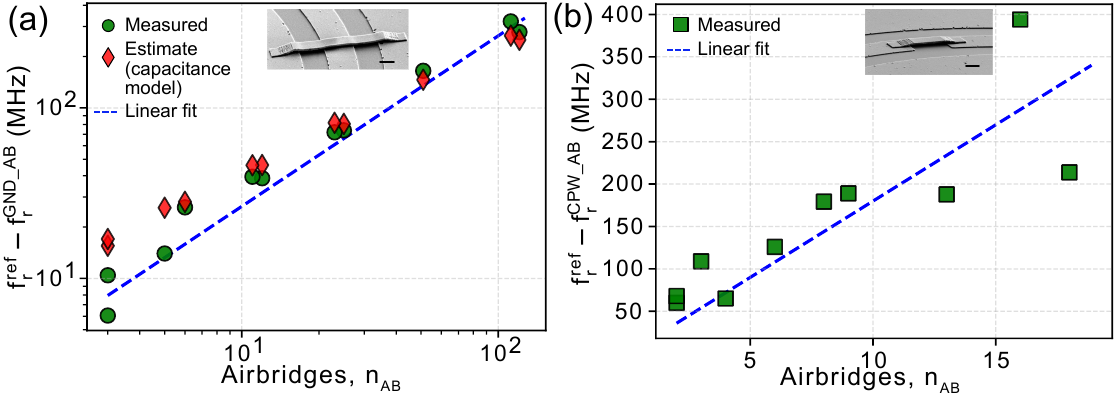}
	\caption{(a) Measured frequency shift (green circle) of resonators on the \emph{Ground-airbridge} chip relative to resonators with the same index on the \emph{reference} chip, along with a linear fit (blue dashed line). The estimated frequency shift based on the airbridge capacitance model shown by red diamonds. (b) Measured frequency shift (green square) of resonators on the \emph{CPW-airbridge} chip relative to resonators with the same index on the \emph{reference} chip, along with a linear fit (blue dashed line). For both panels, the uncertainty in the measured frequency shifts is negligible on the scale of the plot. 
    }
	\label{fig:fig14}
\end{figure}

We carried out the measurement of the resonator chips in \emph{Cryogenic Measurement Setup 1} shown in Fig.\,\ref{fig:fig11}(a) consisting of a Qinu dilution refrigerator and a Vector Network Analyzer from Rohde $\&$  Schwarz for spectroscopy. 
We acquire the complex transmission coefficient $S_{21}$ of the resonators at different intra-resonator photon numbers by varying the input power $P_{in}$. We fit the measured $S_{21}$ data with a phenomenological notch-type resonator model\,\cite{Probst2015a}, 
\begin{equation}
\begin{aligned}
S_{21}(f)
&=
A
\left[
1-
\frac{\left(Q_l /Q_e\right)e^{i\phi} }
{1 + 2 i Q_l \left(f/f_r - 1\right)}
\right]
\end{aligned}
\label{eq:notch_res}
\end{equation}
where $Q_e $ is the external quality factor, $\phi$ accounts for the phase delay between the driving field and the re-emitted intra-resonator cavity field,  $Q_l$ is the loaded quality factor, $f_r$ is the resonance frequency and $A$ denotes the contribution from background transmission. 
To extract the internal quality factor $Q_{i}$, we first estimate the background transmission $A$ for each resonator by fitting the measured $S_{21}$ response acquired at a $P_{\mathrm{in}}$ for which the signal-to-noise ratio is sufficiently high. 
For each resonator, we use the corresponding $A$ to normalize all measured $S_{21}$ responses acquired at different $P_{\mathrm{in}}$, such that $S_{21}^{\mathrm{norm}} =S_{21}/A$. We then fit $S_{21}^{\mathrm{norm}}$ to the resonator model in Eq.\,\ref{eq:notch_res} using the circle-fit method\,\cite{Probst2015a} to estimate $Q_l$, see Fig.\,\ref{fig:fig12}. We subsequently obtain $Q_{i}$ from $Q_l$ using the relation 
\begin{equation}
\begin{aligned}
\frac{1}{Q_{i}}
&=
\frac{1}{Q_l}
-
\frac{\cos\phi}{Q_{e}}
\end{aligned}
\label{eq:Qint}
\end{equation} 
We further extract the average photon number in the CPW resonator, $\left<n_{ph}\right>$ using the relation\,\cite{Bruno2015}, 
\begin{equation}
    \begin{aligned}
        \left<n_{ph}\right> \approx \frac{Q_l^2 \cdot P_{in}}{Q_{e} \cdot 2\pi^2\hbar \cdot f_r^2} 
    \end{aligned}
    \label{eq:photon_num}
\end{equation}
We estimate a total attenuation of $\sim85\,\mathrm{dB}$ at $5\,\mathrm{GHz}$ from all the components within the input line and use the estimated values for all the frequencies while calculating the photon number. Using this relation, we estimate that the photon number in the study ranges from $0.07$ to approximately $10^6$.
\\\\
For the \emph{reference} chip we observe a typical power dependence of the $S_{21}^{\mathrm{norm}}$ data as shown in Fig.\,\ref{fig:fig13}(a). The photon number dependence of the extracted $Q_{i}$ for all the $12$ resonators in Fig.\,\ref{fig:fig13}(b) shows saturation at high-power regime consistent with the saturation of two-level systems within and is typical for a CPW resonator\,\cite{Megrant2012a}.
We observe that the power dependence of  $S_{21}^{\mathrm{norm}}$ data for \emph{Ground-airbridge} chip is similar to the \emph{reference} as shown in Fig.\,\ref{fig:fig13}(c). The photon number dependence of $Q_{i}$ shown in Fig.\,\ref{fig:fig13}(d) highlights the presence of outliers corresponding to $\mathrm{n}_\mathrm{AB}\ge100$ as shown in Fig.\,\ref{fig:fig3}(d), which we attribute to a possible presence of residual resist beneath the airbridges as mentioned in Sec.\,\ref{sec:resonator_spectroscopy}. 
\\\\
Resonators with the same index (shown in Fig.\,\ref{fig:fig3}(a)) have the same length on all three chips. Any difference in the resonance frequencies between the resonator with same index on different chips must therefore stem from the presence of the airbridges. For the \emph{Ground-airbridge} chip, we observe a consistent red-shift in comparison to the \emph{reference} chip, which increases proportionally with the number of airbridges $\mathrm{n}_{\mathrm{AB}}$,  and a linear fit closely follows the measured data, as shown in Fig.\,\ref{fig:fig14}(a). This behavior is consistent with earlier reports\,\cite{chen2014d}, where the airbridges over CPW line acts as shunt capacitors, increasing the effective capacitance and thereby reducing $f_r$. To model this effect, we use the perturbative frequency-shift relation
\begin{equation}
\label{eq:airbridge_frequency_shift_capacitive_estimate}
\begin{aligned}
\frac{\Delta f}{f_r^{\mathrm{ref}}}
\approx
\frac{1}{2}
&\left(
\frac{\Delta C}{C^{\mathrm{ref}}}
\right),\\[4pt]
\Delta f
= f_r^{\mathrm{ref}} - f_r^{\mathrm{GND\_AB}}\quad; \quad&
 \Delta C= C^{\mathrm{GND\_AB}} - C^{\mathrm{ref}}.
\end{aligned}
\end{equation}
Here $f\mathrm{_r^{ref}}$  is the resonance frequency of a resonator on the \emph{reference} chip measured at the lowest driving power and $C\mathrm{^{ref}}=c \cdot l_{\mathrm{res}}/2$ is the total capacitance of the corresponding resonator, where c is the capacitance per unit length and $l_{\mathrm{res}}$ is the resonator length. The quantities with superscript $\mathrm{GND\_AB}$ refer to the resonator of same index on \emph{Ground-airbridge} chip. 
Since the airbridges are distributed at different positions along the quarter-wave resonators, the effective capacitance contribution from each airbridge depends on its positions relative to the voltage node at the grounded end of the resonator. We therefore write   
\begin{equation}
    \begin{aligned}
        \Delta C \approx  \sum_{i = 1}^{\mathrm{n}_{\mathrm{AB}}}C_{\mathrm{AB}} \cdot \sin^2{\left(\frac{\pi x_i}{2l_{\mathrm{res}}}\right)} 
    \end{aligned}
    \label{eq:airbridge_effective_capacitance}
\end{equation}
where $C_{\mathrm{AB}}$ is the capacitance of a single airbridge, $x_i$  is the distance between the $i$-th airbridge and the grounded end of the resonator, and $l_{\mathrm{res}}$ is the resonator length. We estimate $C\mathrm{_{AB}}$ as $\epsilon_0\left(w \cdot l_{\mathrm{AB}}/d\right)$ where  $\epsilon_0$ is the vacuum permittivity, $w\approx4\,\mu\mathrm{m}$ is the bridge width, $l_{\mathrm{AB}}\approx10\,\mu\mathrm{m}$ is the bridge length, $d\approx800\,\mathrm{nm}$ is the bridge height above CPW line. The frequency shift estimated using  this model agrees well with the measured values, as shown in Fig.\,\ref{fig:fig14}(a) , indicating a good agreement with the capacitive loading model. The larger discrepancy at $n_{\mathrm{AB}} < 10$ indicates that, in this lower range, the frequency shift is not determined solely by the additional capacitance contribution from the airbridges. As a result the model in Eq.\,\ref{eq:airbridge_frequency_shift_capacitive_estimate} overestimates the frequency shift for small $n_{\mathrm{AB}}$.
\\\\
In the \emph{CPW-airbridge} chip, the resonators exhibit non-linear behavior, indicated by the bifurcation of the $S_{21}$ response in the high-power regime. This is highlighted by the gray shaded region in Fig.\,\ref{fig:fig13}(e).  In this regime, the linear model described by Eq.\,\ref{eq:notch_res} breaks down and the linear model fit does no longer result in reliable values for  $Q_{i}$ in the high power regime as shown in Fig.\,\ref{fig:fig13}(f). While extensions to nonlinear models exist \cite{Frasca2023a}, here, for simplicity and because we are primarily interested in the low-power $Q_{i}$ factor we  focus on the regime in  which the linear model is still valid.
The measured frequency red-shift for this chip, shown in Fig.\,\ref{fig:fig14}(b) exhibits a dependence on  $n_{\mathrm{AB}}$ similar to the \emph{Ground-airbridge} chip, but with a weaker linear trend and deviates noticeably from the fit for $n_{\mathrm{AB}} > 10$.  We also estimate the Kerr nonlinearity from the measured resonator response to be on the order of a few hundred $\mathrm{Hz}/\mathrm{photon}$, which is much larger than values typically reported for kinetic-inductance-based superconducting resonators, usually in the range of tens of $\mathrm{Hz}/\mathrm{photon}$ \,\cite{Gupta2025b, Khorramshahi2025, Kirsh2021, Frasca2023a}. Both the frequency shift and the non-linearity are therefore unlikely to originate solely from the geometric or kinetic inductance contributions of the airbridges. This leaves the physical origin of non-linear effect as an open question for future investigations. 

\section{Transmon measurement}
\label{sec:supplemental_transmon}

\begin{table}[b]
	\centering  \begin{tabular}{l c}\hline
		Parameters & Value\\
		\hline\\
		Qubit frequency, $\omega_q/2\pi$ & $4.3 - 5.3$\,$\mathrm{GHz}$ \\
		Readout resonator frequency, $\omega_r/2\pi$ & $6.0-7.0$\,$\mathrm{GHz}$ \\
		Charging energy, $E_c/2\pi$ & $208$\,$\mathrm{MHz}$\\
		Effective dispersive shift, $\chi_{\mathrm{eff}}/2\pi$ & $0.3-0.35$\,$\mathrm{MHz}$\\
		Qubit-Readout coupling, $g_{\mathrm{rq}}/2\pi$  &  $55$\,$\mathrm{MHz}$\\
		Resonator-Purcell filter coupling, $g_{\mathrm{rp}}/2\pi$ & $60$\,$\mathrm{MHz}$\\
		Effective decay rate, $\kappa_{\mathrm{eff}}/2\pi$ & $0.6-0.7$\,$\mathrm{MHz}$\\
		\hline
	\end{tabular}
	\caption{Measured transmon device parameters. The parameters are extracted using standard fitting methods\,\cite{Naghiloo2019}.}
	\label{tab:transmon_parameter}
\end{table}

\begin{figure}[t]
\centering
    \includegraphics[width=\linewidth]{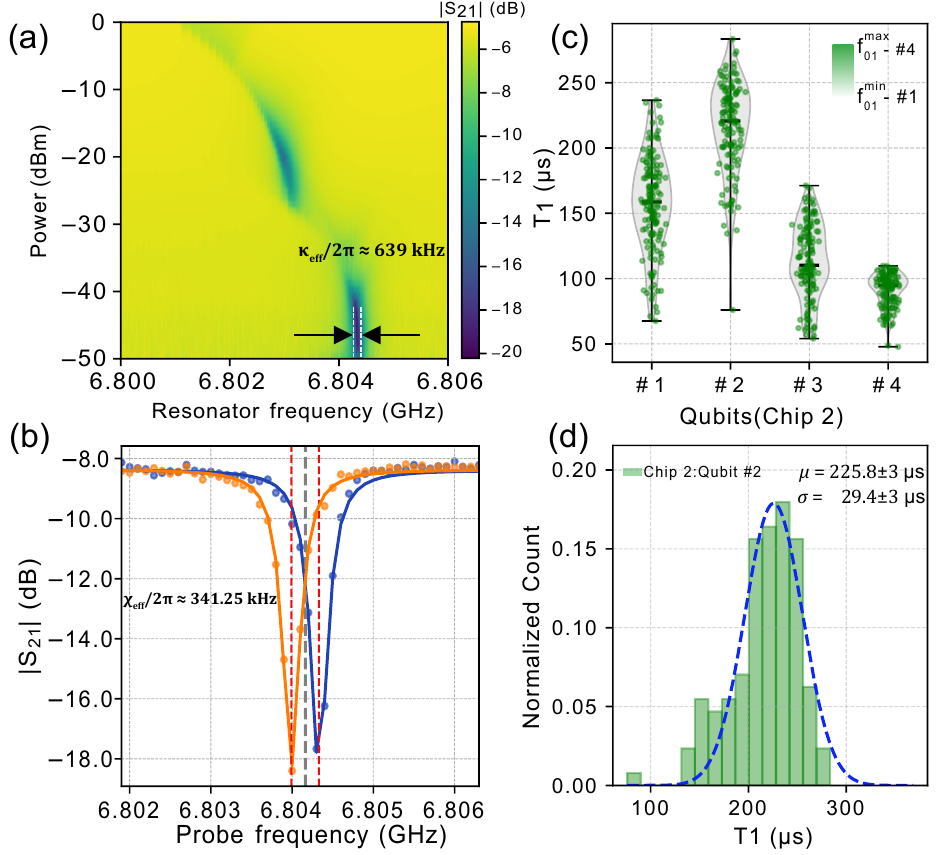}
	\caption{(a) Measured  transmission spectra of  the readout resonator for one of the qubits during one-tone spectroscopy. The arrow shows the region of the spectra $|S_{21}|$ where we perform a Lorentzian fit to extract $\kappa_{\mathrm{eff}}/2\pi$. (b) Measured transmission spectra of the readout line with the qubit in its ground state (blue) and excites state (orange). The solid lines are Lorentzian fits to the data to estimate resonance frequency, $f_0$. (c) Violin plot of the measured $T_1$ values for each transmon on \emph{Chip 2} obtained from 100 repeated measurements. (d) Histogram of the measured $T_1$ values for \emph{Qubit 2} in panel (c), with a Gaussian fit (blue dashed line).}
	\label{fig:fig15}
\end{figure}
We carried out the qubit measurements in two cryogenic setups shown in Fig.\,\ref{fig:fig11}. \emph{Cryogenic Measurement Setup 1} [Fig.\,\ref{fig:fig11}(a)] consists of a dilution refrigerator from Qinu and control electronics unit from QBlox for qubit control and readout. \emph{Cryogenic Measurement Setup 2} [Fig.\,\ref{fig:fig11}(b)] consists of a Bluefors XLD dilution refrigerator and an OPX1000 from Quantum Machines as the room-temperature control electronics unit.  On chip, we perform both control and readout of the transmons through the dedicated resonator-Purcell filter pair it is coupled to. 
\\\\
We characterize the qubits at millikelvin temperature using heterodyne measurements. We first perform one-tone spectroscopy to identify the readout resonance frequency. From the one-tone spectra shown in Fig.\,\ref{fig:fig15}(a), we determine the resonator drive power for readout, which is  $-45\,\mathrm{dBm}$ in this case. At this power, we estimate the effective decay rate $\kappa_{\mathrm{eff}}/2\pi$ from the spectra as shown in Fig.\,\ref{fig:fig15}(a). A Lorentzian fit yields  $\kappa_{\mathrm{eff}}/2\pi$ to be $\sim640\,\mathrm{kHz}$. 
We then perform two tone spectroscopy to determine the qubit transition frequency.  The effective dispersive shift $\chi_{\mathrm{eff}}/2\pi$ is extracted from the resonator response with the qubit prepared in the ground and first excited states, as shown in Fig.\,\ref{fig:fig15}(b). Lorentzian fits to the corresponding spectra yield the dressed readout resonance frequency for the two qubit states. From the frequency difference, we estimate $\chi_{\mathrm{eff}}/2\pi$ to be $\sim340\,\mathrm{kHz}$ which is in good agreement with the designed value. Finally, we calibrate the control and readout signal through Rabi oscillation and single-shot readout before proceeding to the $T_1$ measurement. 
\\\\
For each of the four transmons on the chip we performed over $100$ independent measurements to obtain the $T_1$ distribution as shown in Fig.\,\ref{fig:fig14}(c). We fit a Gaussian distribution to the histogram of $T_1$ distribution shown in Fig.\,\ref{fig:fig14}(d) to obtain the median $T_1$ value for the transmon. We perform similar analysis on all the transmons on three chips presented in Fig.\,\ref{fig:fig4}(d) of Sec.\,\ref{sec: qubit_spectroscopy}

\end{appendix}

\phantomsection\label{LastBibItem}

\end{document}